%% file: main.tex
\documentclass[conference]{IEEEtran}
\IEEEoverridecommandlockouts

\input{00_decl}

\begin{document}

\title{Enhancing TTS Stability in Hebrew using \\ Discrete Semantic Units
}

\author{\IEEEauthorblockN{Ella Zeldes, Or Tal, Yossi Adi}
\IEEEauthorblockA{The School of Computer Science \& Engineering \\
\textit{The Hebrew University of Jerusalem, Israel}\\ 
\texttt{ella.erez@mail.huji.ac.il}}

\vspace{-10pt}
}

\maketitle

\begin{abstract}
This study introduces a refined approach to \tts generation that significantly enhances sampling stability across languages, with a particular focus on Hebrew. By leveraging discrete semantic units with higher phonetic correlation obtained from a self-supervised model, our method addresses the inherent instability often encountered in \tts systems, especially those dealing with non-diacriticized scripts like Hebrew. Utilizing HuBERT codes, our model generates discrete representations that are optimized for \tts tasks, thereby reducing the dependency on diacritic-based text processing. This advancement not only simplifies the language modeling process but also improves the robustness and shows controllability of the speech output due to disentenglement properties of the semantic units. The inclusion of a speaker embedding in the vocoder further aids in capturing the unique vocal characteristics of the speaker, contributing to the naturalness of the synthesized speech. Our experimental results demonstrate that this approach not only maintains high performance in Hebrew but also shows adaptability to English, underscoring its effectiveness in enhancing stability in \tts systems universally. Our method, named \mthd (Language of The Hebrew Man), outperforms existing methods in terms of stability while achieving naturalness and speaker similarity on par with previous methods, making it a compelling choice for future speech synthesis applications. Samples can be found in our page \href{https://pages.cs.huji.ac.il/adiyoss-lab/LoTHM}{pages.cs.huji.ac.il/adiyoss-lab/LoTHM}. 
\end{abstract}


\input{01_intro}
\input{02_related}

\input{03_method}
\input{04_experimental_setup}

\input{05_results}

\input{06_conclusion}

\bibliographystyle{plain}  
\bibliography{refs}  
\end{document}

%% file: 00_decl.tex
\usepackage{etoolbox}
\usepackage{amsfonts}
\usepackage{amssymb}
\usepackage{pifont}
\usepackage[T1]{fontenc} 
\usepackage[utf8]{inputenc} 
\usepackage{amsmath,cite,url}
\usepackage{graphicx}
\usepackage{color}
\usepackage{booktabs}
\usepackage{lineno}
\usepackage{xspace}
\usepackage{bbm}
\usepackage{amsopn}
\usepackage{tabularx}
\usepackage{cuted}
\usepackage{xspace}
\usepackage{hyperref}

\input{definitions}

\newcommand{\newpara}[1]{\vspace{0.02cm}\noindent \textbf{#1}}

\newtoggle{release}
\togglefalse{release}

\iftoggle{release}{
\newcommand{\adios}[1]{}
\newcommand{\yair}[1]{}
\newcommand{\ort}[1]{}
}
{
\newcommand{\adios}[1]{{\color{magenta}Y: #1}}

\newcommand{\ort}[1]{{\color{red}O: #1}}
}

\newcommand{\hub}{HuBERT }
\newcommand{\enc}{EnCodec }
\newcommand{\valle}{VALL-E }

\newcommand{\zac}{{z_{ac}}}
\newcommand{\ztx}{{z_{txt}}}
\newif\ifLM \LMtrue \newcommand{\lm}{\ifLM language model (LM) \LMfalse\else LM \fi}
\newif\ifAR \ARtrue \newcommand{\ar}{\ifAR Auto-Regressive (AR) \ARfalse\else AR \fi}
\newif\ifTTS \TTStrue \newcommand{\tts}{\ifTTS Text-to-Speech (TTS) \TTSfalse\else TTS \fi}
\newif\ifMPD \MPDtrue \newcommand{\mpd}{\ifMPD Multi-Period Discriminator (MPD) \MPDfalse\else MPD \fi}
\newif\ifMSD \MSDtrue \newcommand{\msd}{\ifMSD Multi-Scale Discriminator (MSD) \MSDfalse\else MSD \fi}

\newif\ifWER \WERtrue \newcommand{\wer}{\ifWER Word Error Rate (WER) \WERfalse\else WER \fi}
\newif\ifCER \CERtrue \newcommand{\cer}{\ifCER Character Error Rate (CER)\CERfalse\else CER \fi}
\newif\ifMMS \MMStrue \newcommand{\mms}{\ifMMS Massively Multilingual Speech (MMS) \MMSfalse\else MMS \fi}

\newif\ifASR \ASRtrue \newcommand{\asr}{\ifASR Automatic Speech Recognition (ASR) \ASRfalse\else ASR \fi}

\newif\ifMETHOD \METHODtrue 

\newcommand{\mthd}{\textsc{LoTHM}\xspace}

\usepackage{cite}
\usepackage{amsmath,amssymb,amsfonts}
\usepackage{algorithmic}
\usepackage{graphicx}
\usepackage{textcomp}
\usepackage{xcolor}
\def\BibTeX{{\rm B\kern-.05em{\sc i\kern-.025em b}\kern-.08em
    T\kern-.1667em\lower.7ex\hbox{E}\kern-.125emX}}

%% file: definitions.tex
\usepackage{bm}
\usepackage{multirow}

               \newcommand{\yh}{\hat{y}}



%% file: 01_intro.tex
\section{Introduction}
\label{sec:intro}

Vowel pointing is the inserting of signs used to indicate vowels in certain alphabets. In practice, this is done by adding small marks to letters in scripts that predominantly represent consonants to indicate vowel sounds. This practice is common in languages like Hebrew and Arabic, where the alphabet does not explicitly represent vowels. These diacritical marks are crucial for clarifying pronunciation and distinguishing between words that might otherwise be ambiguous in writing.
However, more often than not, Hebrew speakers read and write without these diacritical marks, relying on their familiarity with the language and context to infer the correct pronunciation. Most available Hebrew text data and \asr systems also produce transcriptions without diacritics, presenting significant challenges for \tts systems in Hebrew.
Previous research in Hebrew \tts~\cite{sharoni2023saspeech, pratap2024scaling} has typically involved adding diacritics to text using tools like Nakdan~\cite{shmidman2020nakdan} before feeding it into the \tts system.
Other approaches~\cite{roth2024language} have experimented with generating \tts output directly from non-diacriticized text using acoustic tokenizers such as \enc \cite{defossez2022high} to discretize the audio and train a \lm on top of the discrete token stream.
These type of acoustic tokenizers are trained to minimize a reconstruction loss rather than maximizing mutual information across latent vectors like~ \cite{baevski2020wav2vec,hsu2021hubert}.
\begin{figure}[t!]
\centering
\includegraphics[width=0.8\linewidth ]{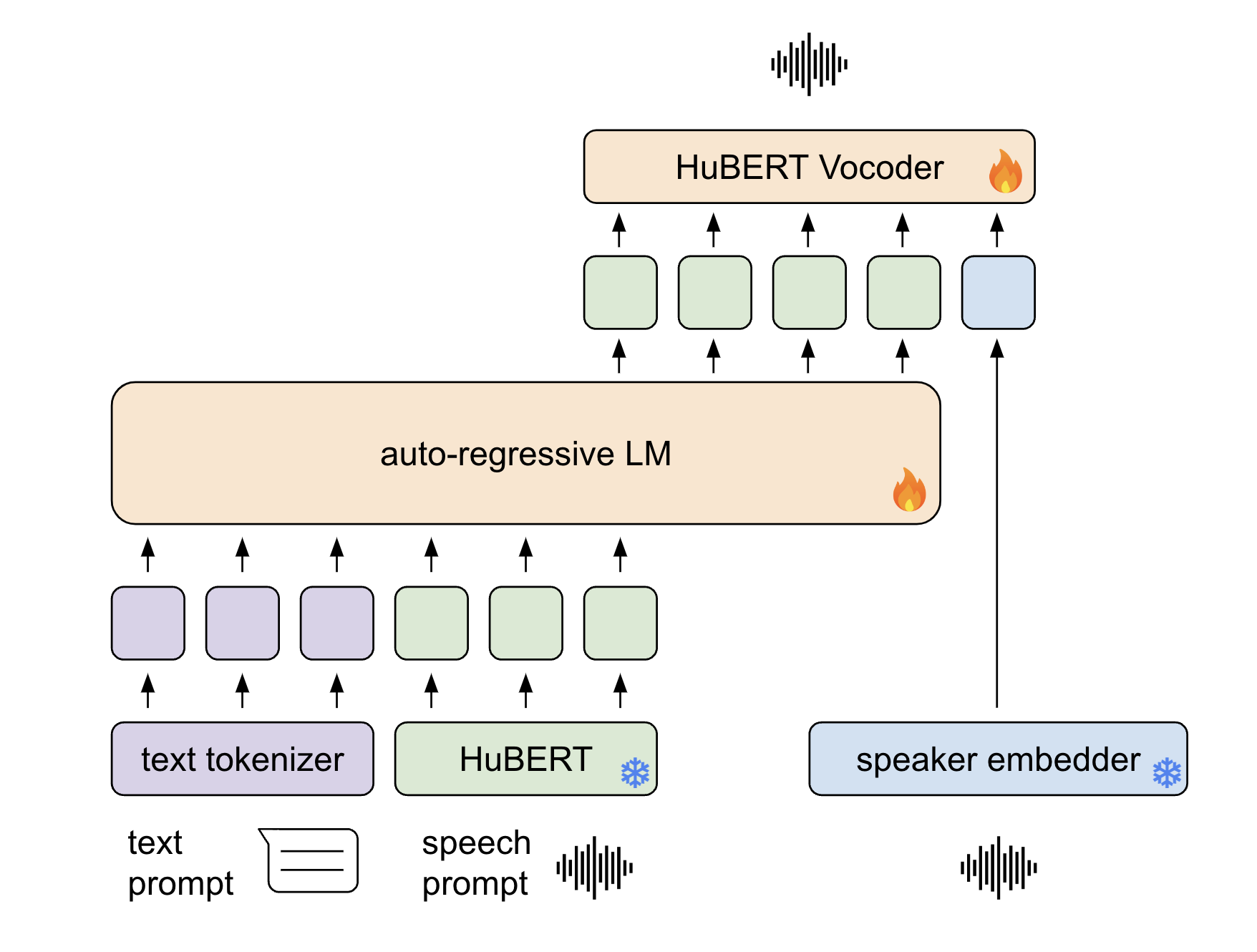}
\caption{A high level overview of the propose modeling architecture. \label{fig:arch}}
\vspace{-5pt}
\end{figure}
Semantic tokens, such as \hub tokens, are designed to maximize mutual information across vectors, while acoustic tokens, like \enc tokens, are optimized to minimize reconstruction-based metrics. Previous studies have demonstrated that semantic tokens exhibit a significantly higher correlation with phonetic content~\cite{sicherman2023analysing}, leading to improved \asr performance~\cite{mousavi2024dasb} and enhanced \tts performance~\cite{ye2024codec, mousavi2024dasb} compared to acoustic tokens.
Considering the \tts setup, where we could decipher the phonetic content from the text condition, using such acoustic tokenizers could benefit the naturalness and diversity of the generated speech~\cite{wang2023neural}.
The usage of acoustic tokenizers in \tts systems usually involves \emph{best-of-k sampling}, where k samples are generated and the one with the best \wer is chosen, to improve content match~\cite{chen2024vall, roth2024language}.
The need to sample a few times arises from the instability nature of the model, where different samples vary in quality. 
In this study, we employ \hub tokens to achieve more stable and consistent outcomes in \tts systems.
Our method not only simplifies language modeling by leveraging the high mutual information within the units but also avoids the complexity of modeling multiple discrete token streams. Previous approaches, like VALL-E~\cite{chen2024vall}, utilize an autoregressive model to generate the first EnCodec codebook, followed by a non-autoregressive model for subsequent codebooks. Similarly, SpearTTS~\cite{kharitonov2023speak} employs two separate models: one for extracting semantic tokens and another for generating acoustic tokens. In contrast, our approach consolidates these tasks into a single language model, which not only streamlines the process but also enhances stability and robustness by reducing the risk of error propagation inherent in cascade models. We then employ a unit-based vocoder, similar to the methods described in~\cite{polyak2021speech, kreuk2021textless, lee2021direct} to synthesize a time-domain signal. Following such an approach further improving the efficiency and overall performance.
To enhance the model's ability to capture the speaker's unique vocal characteristics, we incorporate a speaker embedding vector into the unit-based vocoder. This approach, inspired by techniques used in speech resynthesis~\cite{polyak2021speech}, emotion conversion~\cite{kreuk2021textless}, and speech-to-speech translation~\cite{kim2024transentence}, contributes to producing a more natural and expressive speech output.
By deriving \hub codes directly from audio data, we utilize a much larger and  diverse dataset compared to traditional diacritization based methods, enabling our model to learn more nuanced patterns and improve overall performance.
Our main contributions are: (i) We present \mthd , an LM approach for non-diacritized Hebrew using semantic tokens; (ii) We demonstrate that following such an approach results in more robust and stable sampling. We demonstrate it on both Hebrew and English \tts. (iii) We show that following such an approach allows better controllability of the generated speech, particularly in disentangling speaker identity and speaking rate.

%% file: 02_related.tex
\section{Related Work}
\label{sec:rel}
\newpara{Content Match and Stability.}
\tts has been a highly researched topic going back a long way before the introduction of Deep Neural Networks (DNN), utilizing classic speech processing algorithms to synthesize speech, e.g~\cite{griffin1984signal}. 
Recent development have reached impressive performance in terms of content match, rapidly closing the gap on human performance in English~\cite{wang2023neural,chen2024vall}.
These type of Text-based approaches usually lean on acoustic representation like EnCodec~\cite{defossez2022high}, SoundStream~\cite{zeghidour2021soundstream}
which results in an inherent caveat in sampling stability \cite{chen2024vall}.
Several studies have demonstrated that incorporating phonetic-correlated latent codes enhances content matching across various speech-related applications~\cite{tal2022systematic, nguyen2024spirit, kim2023transduce}. For instance, \cite{kim2023transduce} utilizes a transducer-based system to generate semantic tokens, resulting in improved and more stable performance compared to VALL-E~\cite{wang2023neural}. 
Textless speech generation approaches have tried generating coherent speech, yet results still remain a far fetch from text conditioned approaches.
To successfully generate coherent speech, these type of textless models usually require incorporating usage of more phonetic-related latent codes~\cite{lakhotia2021generative,nguyen2024spirit} or start from pretrained text generation models~\cite{hassid2024textually}.
In \cite{pratap2024scaling}, the authors aim to develop various speech technologies such as \asr, \tts, and Language ID across over 1,000 languages. For Hebrew, their TTS approach involves adding diacritics before predicting representations derived from the pre-trained multilingual self-supervised model, wav2vec \cite{baevski2020wav2vec}, which similarly to \hub is a more phonetic-oriented latent representation.

\newpara{Controllable Speech Generation}
Disentangling various aspects of speech has been explored in prior work \cite{kim2023transduce, polyak2021speech, nguyen2023expresso}. For example, \cite{kim2023transduce} demonstrate that their transducer-based system effectively controls prosody and speaking speed in generated speech. Similarly, speech re-synthesis~\cite{polyak2021speech} has shown the ability to manipulate speech synthesis by controlling factors such as speech content, speaker identity, and pitch.

\newpara{Hebrew TTS.}
The model introduced in SASPEECH~\cite{sharoni2023saspeech} is a single speaker Hebrew \tts model, combining a neural Hidden Markov Model (HMM) with normalizing flows, utilizing HiFi-GAN neural vocoder for speech synthesis. Similar to MMS, the model relies on diacritic prediction from a separate module, Nakdimon\cite{shmidman2020nakdan}, which is trained on Hebrew text with diacritics, posing challenges for scaling to large datasets, as the model requires accurate diacritic prediction.
A recent approach that bypasses the need for diacritic-based text processing is HebTTS\cite{roth2024language}. This method infers diacritic-free text into EnCodec tokens given a 3-second audio prompt along with a text prompt. While this approach demonstrates strong results in terms of naturalness, it requires multiple sampling attempts to achieve satisfactory results in terms of content accuracy.

%% file: 03_method.tex
\section{Method}
\label{sec:method}

\newpara{Problem Setting.}
Given an input audio prompt $x_r\in\mathbb{R}^{f_s \cdot d_p}$ of duration $d_p$ [sec] sampled at $f_s$ Hz, a text condition $c$ and a another input audio $x_{s}$ of possibly a different speaker, our goal is to generate an audio $\yh$ that correlates to the speaker in $x_s$ and the speaking rate of $x_r$ while matching the given text $c$.  To accomplish that, we tokenize input audio $x_r$ and text prompts $c$ using acoustic and text tokenizers, and embed  $x_s$ into a fixed shape representation $s\in\mathbb{R}^{D_{sp}}$, using a pre-trained speaker embedder. 
Specifically, the audio tokenizer maps the given input prompt $x_r$ to a discrete sequence $z_{ac}\in\{N_{ac}\}^{f_r\cdot d_p}$  where $N_{ac}$ is the vocabulary size and $f_r$ is the tokenizer's frame rate with respect to the temporal axis.
Similarly, the text tokenizer maps the given textual prompt $c$ to a discrete sequence $z_{txt}\in \{N_{txt}\}^{T}$ where ${N_{txt}}$ is the number of possible tokens and $T$ is the max length of tokens. 
Our goal is to learn $f_{\theta}(\zac,\ztx, s)=\yh$ that maps the given inputs to an estimated waveform $\yh\in\mathbb{R}^{f_s \cdot d_g}$, $d_g$ being the generated duration in seconds, which corresponds to the textual content in $c$, the speaker in $x_s$ and maintain the speaking rate of the audio prompt $x_r$. 
Specifically, during training, given a ground truth sample $y$ and its matching transcription $c$, we define $x_r$ to be the first $d_p$ seconds of $y$, and $x_s$ to be $x_r$.
We could further decompose $f_\theta=(g\circ h)_\theta$ to two independent modules: (i) a language model $h(\zac, \ztx)=q\in\mathbb{R}^{f_r \cdot d_g}$ and (ii) a vocoder $g(q, s)=\yh\in\mathbb{R}^{f_s \cdot d_g}$, where we train each module separately and cascade them during inference.

\newpara{Acoustic and text tokenizers.}
We use a pretrained model released by ~\cite{nguyen2023expresso}, who trained a \hub model~\cite{hsu2021hubert} under a multi-lingual setup, as well as its computed K-means clustering ($2000$ centroids).
In order to obtain the discrete acoustic token sequence $\zac$, we first extract the $12^{\text{th}}$ transformer layer's output of the pretrained model. Then, we discretize the extracted latent by replacing each latent vector with its corresponding closest centroid index in the mentioned K-means clustering.
As for the text tokenizer, we use the pre-trained word-piece text tokenizer of AlephBERT \cite{seker2021alephbert} with vocabulary size of $52$K tokens.

\newpara{Speaker embedding and speech vocoder.}
To synthesize a time-domain signal, we use a unit-based vocoder similarly to~\cite{polyak2021speech}. By design, \hub tokens are phoneme-oriented and known to discard some acoustic information, e.g speaker identity~\cite{kharitonov2022textless}. We therefore feed a speaker embedding representation as an additional input to the vocoder, hence enabling the system to capture unique vocal characteristics, while not being limited to a predefined set of seen speakers. We train the vocoder from scratch on both English and Hebrew data while conditioning the vocoder on the input domain $\in$ \{Heb,En\} by using a binary learned embedding. To obtain a speaker embedding, we use a pretrained speaker verification model~\cite{desplanques2020ecapa} which receives an audio prompt as input and outputs a single speaker vector. Overall, we concatenate the discrete \hub units sequences with a speaker embedding vector and a binary domain embedding as input to the vocoder which outputs an estimated waveform $\yh$.
The neural vocoder architecture is a modified version of HiFi-GAN \cite{kong2020hifi} comprised of a generator, \mpd and \msd. 

\newpara{Language Model.}
Inspired by recent language models based approaches for \tts \cite{wang2023neural}, our model is an \ar \lm using a similar modeling to the \ar part of \valle \cite{wang2023neural}. Formally, the \lm $h$ receives $\zac,\ztx$ as input, embeds them, adds a sinusoidal positional embedding to each of them and concatenates the two with a separator token in between.
During inference, we use a top-$p$ sampling strategy, according to which - at time step $t$ we sample $q_t\in\{N_{ac}\}$ from the top-$p$ probability mass, w.r.t its corresponding distribution.

\newpara{Training objective.}
During the \lm training we employ a standard teacher forcing paradigm where we minimize the cross entropy objective for next token prediction.
For the vocoder training, the core training objective combines three main components: adversarial loss, feature matching loss, which is the L1 distance between the intermediate feature representations produced by the discriminators, and a mel-spectrogram loss (minimizing the L1 distance between the mel-spectrograms of the real and generated audio)~\cite{kong2020hifi}.

%% file: 04_experimental_setup.tex
\section{Experimental setup}
\label{sec:exp_setup}

\newpara{Data.}
We utilized the HebDB dataset\cite{turetzky2024hebdb} and ivrit.ai dataset\cite{marmor2023ivrit}, which together consist of approximately $4500$ hours of data. These datasets were obtained from local podcasts that included spontaneous dialogues with multiple speakers discussing a variety of topics (e.g., economy, politics, sports, etc.). We preprocess the data in a similar way to~\cite{roth2024language}. 
Since the Hebrew dataset consists mainly of non-sterile data, to better train the vocoder to achieve clearer voice, we additionally use the VCTK dataset\cite{veaux2013voice} ($\sim 44$ Hr, $109$ speakers, English) making it a multi-lingual vocoder.
To train the English \lm we used LibriTTS~\cite{zen2019libritts} ($585$ Hr, $2456$ speakers). 
For Hebrew evaluation we use the SASPEECH dataset\cite{sharoni2023saspeech} ($30$ Hr, single speaker), and for English we use the test set of LibriTTS ($16$ Hr, 72 speakers).
We downsample the audio to match our used audio tokenizer's expected sampling rate: $16$KHz for \hub, and $24$KHz for \enc.

\begin{table}[t!]
\centering
\caption{TTS results for \mthd and the baseline methods while sampling one time (x1) or three times (x3) from the models. Human study also include 95\% confidence intervals. *indicates statistically significant results, p-value $\leq$ 0.05. }
\label{tab:main}
\resizebox{\columnwidth}{!}{
\begin{tabular}{@{}l|ccc|ccc@{}}
\toprule
\multirow{2}{*}{\textbf{Model}} & \multicolumn{3}{c}{\textbf{Obj. Metrics}} & \multicolumn{3}{c}{\textbf{Human Study}} \\ \cmidrule(l){2-7}
& \textbf{WER$\downarrow$} & \textbf{CER$\downarrow$} & \textbf{Spk. Sim.$\uparrow$} & \textbf{Naturalness$\uparrow$} & \textbf{Content$\uparrow$} & \textbf{Spk. Sim.$\uparrow$} \\
\midrule
\textbf{Reference} & 0.07 & 0.03 & 0.96 & 4.81 $\pm$0.07 & 4.74 $\pm$0.08 & - \\
\midrule
\textbf{MMS} & 0.23 & 0.07 & - & 2.56 $\pm$0.18 & 2.28 $\pm$0.15 & - \\
\textbf{SASPEECH} & 0.20 & 0.08 & 0.81 & 3.57 $\pm$0.18 & 4.16 $\pm$0.14 & 2.74 $\pm$0.21 \\
\textbf{HebTTS (x1)} & 0.34 & 0.18 & \textbf{0.94} & \textbf{3.97} $\pm$0.18 & 4.31 $\pm$0.15 & 2.94 $\pm$0.16 \\
\textbf{HebTTS (x3)} & 0.19 & 0.08 & \textbf{0.94} & - & - & - \\
\midrule
\textbf{\mthd (x1)} & 0.17* & 0.09* & \textbf{0.94} & \textbf{3.97} $\pm$0.17 & \textbf{4.45} $\pm$0.14 & \textbf{3.08} $\pm$0.15 \\
\textbf{\mthd (x3)} & \textbf{0.10}* & \textbf{0.03}* & \textbf{0.94} & - & - & - \\
\bottomrule
\end{tabular}}
\end{table}

\newpara{Implementation details.}
The \ar model is a causal transformer architecture with $12$ layers, $16$ attention heads, an embedding dimension of $1024$, and a feed-forward block dimension of $4096$. We train the model on varying audio length sequences ranging between $1$~to~$18$ seconds. The model is trained using $4$ NVIDIA A$30$ $24$GB GPUs with a total batch size of $8$K acoustic tokens. We optimize the model using EDEN scheduler~\cite{yao2023zipformer} with a starting learning rate of $5\times10^{-2}$. 
We train the vocoder on $1$-$18$ seconds long audio samples using the same GPU setup and a batch size of 16 audio samples. We optimize the model using Adam optimizer and exponential scheduler, with a starting learning rate of $2\times10^{-4}$. Our vocoder is a HiFi-GAN generator, trained with the following discriminator specs: scales=\{$1,2,4$\} and periods=\{$2, 3, 5, 7, 11$\}. see \cite{kong2020hifi} for detailed description.

\newpara{Evaluation metrics.}
We evaluate the results of our model using both objective and subjective (human evaluation) metrics. The objective metrics are: Word Error Rate (WER), \cer, speaker similarity and speaking rate, whereas the subjective metrics are human evaluation of speaker similarity, quality and naturalness of generated audios.

\newpara{WER and CER.}
Given the generated audio, we use Whisper V2-Large \cite{radford2023robust} to transcribe it, and then calculate the \wer and \cer with respect to the ground truth transcription. We estimate these metrics on $100$ randomly sampled text prompts from SASPEECH\cite{sharoni2023saspeech} dataset. We normalize the text by removing all punctuations from both original and transcribed texts. To calibrate the results with the errors produced by Whisper, we also compute \wer and \cer between the input text and transcribed text of the original recording.
To account for sampling stability, as in HebTTS~\cite{roth2024language}, we evaluate \wer and \cer on best-of-$3$ as well as a single sample paradigm.

\newpara{Speaker Similarity and speaking rate.}
We evaluate the speaker similarity by calculating the cosine similarity between the speaker embedding of the generated speech and the speaker embedding of the enrolled 3 seconds speech prompt. 
In order to achieve unbiased results, we use a different vector embedding than the one used to condition our vocoder on. Specifically, we use the xvector embedding\cite{snyder2018spoken}.
During the speaking rate evaluation, we use a fixed speaker embedding, that differs from the speakers used in the audio prompt. To evaluate the speaking rate we first transcribe the generated audio using Whisper V2-Large and obtain the number of generated words. Then, we simply divide the number of words by the duration of the generated audio.

\begin{table}[t!]
\centering
\caption{WER and CER using different tokenizers for speech generation trained on LibriTTS (English). \\ *indicates statistically significant results, p-value $\leq$ 0.05.}
\label{tab:eng}
\resizebox{0.55\columnwidth}{!}{
\begin{tabular}{@{}l|cc@{}}
\toprule
   \textbf{Tokenizer units}     & \textbf{WER$\downarrow$} & \textbf{CER$\downarrow$} \\
\midrule
EnCodec (x1) & 0.22         & 0.13          \\
EnCodec (x3) & 0.13         & 0.08          \\
HuBERT (x1)  & 0.15* & 0.09* \\
HuBERT (x3) & \textbf{0.09}* & \textbf{0.04}* \\
\midrule
GT      & 0.07          & 0.02         \\
\bottomrule
\end{tabular}}
\end{table}

\newpara{Human Study.} 
We handpicked $10$ distinct samples, including samples that could have different pronunciation mapped to the same textual transcription, relying on context to conclude the correct form of speech.
All subjective evaluations are scaled between $1$ to $5$, where $5$ being the best rating.
Raters are asked to rate the naturalness of the generated audios and their fidelity to the text. In addition they are asked to rate the speaker similarity to a reference audio.
All of the raters are native Hebrew speakers, with average of 15 unique raters per sample.

\newpara{Baselines.}
We evaluate the proposed method against three baseline models: (i) \mms \cite{pratap2024scaling}, (ii) SASPEECH model \cite{sharoni2023saspeech}, and (iii) HebTTS \cite{roth2024language}. The \mms model employs a multilingual wav2vec2.0 \cite{baevski2022data2vec} trained on approximately $500k$ hours of data across $1,107$ languages, including $25$ hours in Hebrew. The SASPEECH model utilizes a neural HMM combined with normalizing flows to capture the highly non-Gaussian distribution of acoustic features and is trained on 30 hours of single-speaker, high-quality data from the "Hayot-Kiss" podcast \cite{sharoni2023saspeech}. Both methods rely on external models for predicting diacritics. The HebTTS model is a LM based system similar to \cite{chen2024vall} that uses a pre-trained \enc for both encoding and decoding. For all baselines, we use the official pre-trained models provided by the authors and follow their text pre-processing pipelines.

%% file: 05_results.tex
\section{Results}
\label{sec:results}

\newpara{Text match evaluation.}
Results in Table~\ref{tab:main} draw a comparison of our suggested method and the mentioned baselines. Interestingly, as our \lm and the \lm backbone of HebTTS\cite{roth2024language} are relatively similar and differ mainly in the used audio-tokenizer, it seems that using \hub codes, rather than a more acoustic oriented codes, have improved \wer and \cer by a factor of $2$ on both single and best-of-3 sampling, suggesting that our method offers a more robust sampling process in terms of text-match.
To further explore this observation, we repeat the experiment on English, trained on LibriTTS\cite{zen2019libritts}.
Results drawn in Table~\ref{tab:eng} show a similar trend and further supports the sampling stability mentioned above.
Moreover, since in non-diacritic languages different pronunciations with different meanings can be automatically transcribed to the same word, the importance of subjective evaluation is even more crucial.
The results of the Human study in Table~\ref{tab:main} emphasize a notable improvement in terms of pronunciation correlation seen in the observed content match metric.
In addition to content match, subjective evaluations suggest that our method is at least on-par in terms of generated speech naturalness w.r.t the baselines.

\begin{figure}[t!]
    \centering
    \includegraphics[width=0.8\linewidth]{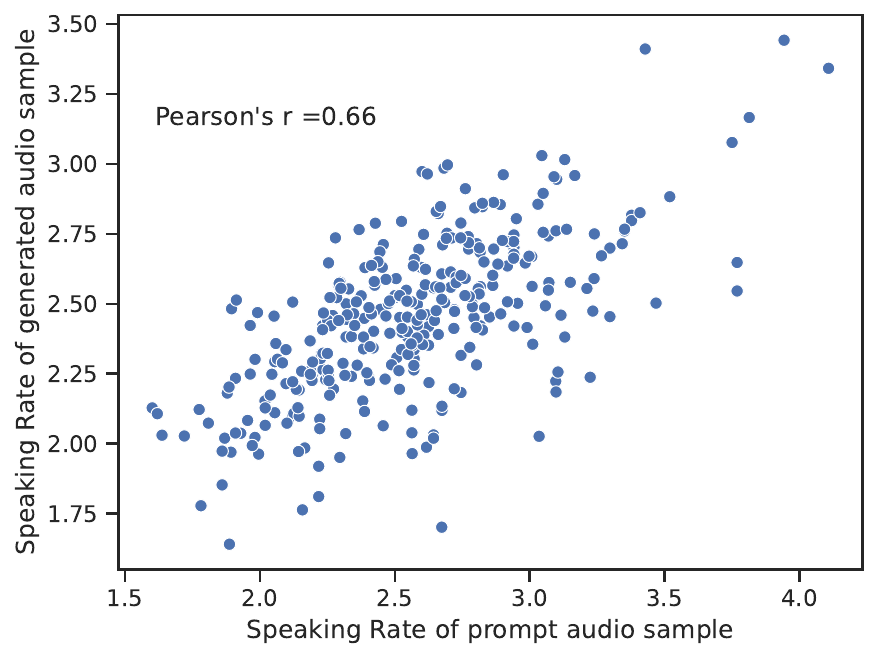}
    \caption{Correlation between the speaking rate of the generated audio and the prompt, measured in words per second, with a strong correlation of 0.66.\label{fig:corr}} 
    \vspace{-0.2cm}
\end{figure}

\newpara{Disentanglement properties.}
By design, we pass a speaker vector to the vocoder after sampling a \hub codes from the LM.
Results in Tab.~\ref{tab:main} suggest that our approach achieves at least comparable performance on subjective results and matching performance on objective results in terms of speaker similarity. Moreover, repeating the experiment while using a fixed unseen speaker vector, differing from the speaker in the given audio prompt, we observe a $0.89$ objective speaker similarity, further emphasizing the speaker disentanglement property of our approach.
To evaluate the speaking rate disentanglement we first partition the test set into $70$ equally spaced bins w.r.t to the speaking rate ($\#\text{words} / \text{sec}$). We then sample at most 10 instances from each bin to be included in this experiment sample set.
Fig.~\ref{fig:corr} draws the correlation between the given prompt and generated audio speaking rates. Generations show $0.66$ correlation w.r.t the ground truth.

%% file: 06_conclusion.tex
\section{Conclusion}
\label{sec:con}
In this work, we present an innovative approach to \tts generation that leverages self-supervised representations with higher phonetic correlation to enhance sampling stability.
Our method, \mthd, shows significant improvements to the stability and controllability of speech synthesis, particularly in handling languages with non-diacriticized scripts such as Hebrew.
Future work could extend this approach for other languages with complex phonetic structures and exploring the integration of emotion and intonation  to further enhance the expressiveness and realism of the synthesized speech.